\documentclass[aps,prl,twocolumn,superscriptaddress,letterpaper,floatfix]{revtex4-1}

\usepackage{epsfig}
\usepackage{graphicx}

\usepackage[centertags,sumlimits,intlimits,namelimits,reqno]{amsmath}
\usepackage{bbm,amsfonts}
\usepackage{amsthm}

\input{Qcircuit}

\theoremstyle{plain}
\newtheorem{theorem}{Theorem}

\newtheorem{fact}[theorem]{Fact}

\newcommand{\Ref}[1]{(\ref{#1})}

\def\Tr{\mathrm{Tr}}

\def\E{{\mathcal E}}

\def\>{\rangle}
\def\<{\langle}

\newcommand{\bbra}[1]{\langle {#1}|}

\newcommand{\Haar}{\mbox{\scriptsize Haar}}
\newcommand{\id}{\leavevmode\hbox{\rm\small1\normalsize\kern-.33em1}}

\def\bfact{\begin{fact}}
\def\efact{\end{fact}}

\def\bv{\left( \begin{matrix}}
\def\ev{\end{matrix} \right)}

\def\dag{^\dagger}

\def\be{\begin{equation}}
\def\ee{\end{equation}}
\def\bes{\begin{eqnarray}}
\def\ees{\end{eqnarray}}
\def\bess{\begin{eqnarray*}}
\def\eess{\end{eqnarray*}}

\def\ii{{\cal I}}

\begin{document}

\title{Exact and Approximate Unitary 2-Designs \\ and their Application to Fidelity Estimation}

\author{Christoph Dankert}
\affiliation{David R.~Cheriton School of Computer Science and
Institute for Quantum Computing, University of Waterloo}
\author{Richard Cleve}
\affiliation{David R.~Cheriton School of Computer Science and
Institute for Quantum Computing, University of Waterloo}
\affiliation{Perimeter Institute for Theoretical Physics, Waterloo}
%\affiliation{School of Computing Science and Institute for Quantum
%Computing, University of Waterloo}
\author{Joseph Emerson}
\affiliation{Department of Applied Mathematics and Institute for
Quantum Computing, University of Waterloo}
\author{Etera Livine}
\affiliation{Perimeter Institute for Theoretical Physics, Waterloo}

%------------------------------------------------------------------------------&
\begin{abstract}
\noindent
We develop the concept of a \emph{unitary $t$-design} as a means of
expressing operationally useful subsets of the stochastic properties
of the uniform (Haar) measure on the unitary group $U(2^n)$ on $n$ qubits.
In particular, sets of unitaries forming $2$-designs have wide
applicability to quantum information protocols. We devise an $O(n)$-size
in-place circuit construction for an \textit{approximate} unitary $2$-design.
We then show that this can be used to construct an efficient protocol
for experimentally characterizing the fidelity of a quantum process on $n$ qubits
with quantum circuits of size $O(n)$ without requiring any ancilla qubits, thereby
improving upon previous approaches.

\end{abstract}

\date{April 27, 2009}

\maketitle

%------------------------------------------------------------------------------&
\section{Introduction}

\noindent
The importance of generating random states and random unitary
operators in quantum information processors has become increasingly
clear from the growing number of algorithms and protocols that
presume such a resource
\cite{AMTdW01,DLT02,Hayden-Bennett-Harrow,AS03,EAZ05,Emerson03,RRS05-Sen05,
ADHW06}. For many algorithms and protocols the invariant (Haar)
measure on the unitary group $U(D)$ is the natural randomization
measure \cite{Hayden-Bennett-Harrow,EAZ05,RRS05-Sen05,ADHW06}. It is
well-known that generating Haar-random unitary operators on a
quantum information processor is inefficient: the number of gates
grows exponentially with the number of qubits. Consequently it is
useful to identify subsets of the unitary group that can adequately
simulate the Haar-measure for a given class of operational tasks and
to identify efficient gate decompositions for these subsets.
%See also \cite{ADHW06} for further discussion examples.

Quantum data hiding \cite{DLT02}, based on a process known as
bilateral twirling \cite{BDSW}, is an example where such a subset
has been identified. Sampling over the discrete subset of $U(2^n)$
known as the Clifford group is sufficient for this
task~\cite{DLT02}, and the number of gates required to implement
such operations is $O(n^2)$ for $n$-qubit systems. It has been shown
recently in~\cite{ADHW06} that a large family of protocols can be
described in terms of bilateral twirling with Haar-random unitary
operators, and noted that the Clifford group is sufficient to
implement them. A related protocol, twirling a quantum channel with
Haar-random unitary operators, has been studied recently
for the purpose of reducing a generic gate to a standard form, which is
among the results of Ref.~\cite{Cirac05}, where it was
also shown that the Clifford group is also an adequate substitute for
that task. Moreover, the task of generating typical generic
entanglement has been studied in Ref.~\cite{ODP06}, where it was shown
that a discrete subset of elements of
the unitary group that are constructible with $O(n^3)$ gates are sufficient
for that purpose. Each of the above tasks satisfies the unitary 2-design condition
and is therefore subsumed by our work.

%In this paper we propose a unified framework for analyzing the degree
%of randomization required by the above and similar protocols. Our
%approach is based on a generalization of the concept of quantum
%state $t$-design \cite{RenesKR05}, and, in the first non-trivial
%case of $t=2$, we give an explicit constructions for the exact and
%approximate implementations of a unitary 2-design. We then discuss
%how our results are useful in the context of experimentally
%estimating the average fidelity of a quantum channel.  In particular
%we show how the(exponential) protocol of Ref.~\cite{EAZ05} can be
%achieved exactly with $O(n^2)$ (and, approximately, with  $O(n
%\log 1/\varepsilon)$) operations for $n$-qubit systems (where
%$\varepsilon$ refers to the probability that the approximate twirl
%fails to simulate the Haar-twirl).

In this paper we propose the concept of a \textit{unitary $t$-design}
as a generalization of the concept of quantum state $t$-design \cite{RenesKR05}, and, in the first non-trivial case of $t=2$, we give explicit efficient constructions for them as quantum circuits.
Our circuits have size $O(n^2)$ for exact implementations and
$O(n \log 1/\varepsilon)$ for approximations within $\varepsilon$ (where
$\varepsilon$ is the degree of closeness to the exact case, which is
formalized in the next section).
We then discuss how our results are useful in the context of experimentally
estimating the average fidelity of a quantum channel.
In particular, we show how the (exponentially inefficient) protocol of Ref.~\cite{EAZ05}
can be achieved efficiently with $O(n)$ operations for $n$-qubit systems.

%\begin{figure}
%\begin{equation*}
%\Qcircuit @C=1.2em @R=1em {
%& \multigate{3}{U^{\dag}} & \multigate{3}{\Lambda} & \multigate{3}{U} & \qw \\
%& \ghost{U^{\dag}} & \ghost{\Lambda} & \ghost{3}{U} & \qw \\
%& \ghost{U^{\dag}} & \ghost{\Lambda} & \ghost{3}{U} & \qw \\
%& \ghost{U^{\dag}} & \ghost{\Lambda} & \ghost{3}{U} & \qw
%}
%\end{equation*}
%\end{figure}

%------------------------------------------------------------------------------&
\section{Definitions and summary of results}

\noindent
We define a \emph{unitary $t$-design} for $D$ dimensions as a finite
set $\{ U_k\}_{k=1}^K \subset \mathcal{U}(D)$ of unitary operators
on $\mathbb{C}^D$ such that,
for every polynomial $P_{(t,t)}(U)$ of degree at most $t$ in the
matrix elements of $U$ and at most $t$ in the complex conjugates
of those matrix elements (i.e., a polynomial of degree at most $(t,t)$),
\be\label{eqn:t-design}
\frac{1}{K} \sum_{k=1}^K P_{(t,t)}(U_k)
= \int_{\mathcal{U}(D)} dU\; P_{(t,t)}(U),
\ee
where, unless otherwise specified, integrals over $\mathcal{U}(D)$ are
with respect to the unitarily invariant Haar measure (for an equivalent
definition in terms of representation theory, see Note~\cite{endnote2}).
This definition is a natural extension of the definition of $t$-designs for quantum states.

The focus of the current paper is on the $t=2$ case; however,
there are specific applications for other values of $t$ (see Note~\cite{endnote3}).
The connection with operational tasks in the case $t=2$ can be seen as follows.
Consider a general quantum channel $\Lambda$ acting on $D$-dimensional
quantum states.
Such a (super)operator is a completely positive trace preserving linear map
acting on $L(\mathbb{C}^D)$, the algebra of linear operators on
$\mathbb{C}^D$.
Suppose that $\Lambda$ is conjugated by a randomly chosen unitary operation
with respect to a probability measure $\mu$ on $\mathcal{U}(D)$.
That is, $U$ is chosen according to $\mu$ and then the channel is modified
to $\hat{U}\dag \circ \Lambda \circ \hat{U}$, where
\begin{eqnarray}
\hat{U}(\rho) & = & U \rho\, U{\dag} \nonumber \\
\hat{U}\dag(\rho) & = & U{\dag} \rho\, U.
\end{eqnarray}
Denoting the resulting superoperator by $\mathbb{E}_{\mu}(\Lambda)$, we have
\begin{equation}
\mathbb{E}_{\mu}(\Lambda) : \rho \mapsto \int_{\mathcal{U}(D)}
d{\mu}(U) \; U\dag \Lambda(U \rho\, U\dag) U
\end{equation}
for all $X \in L(\mathbb{C}^D)$ (including the operationally significant
case where the input to the channel $X$ is a density operator).
The process that transforms any superoperator $\Lambda$ to the
superoperator $\mathbb{E}_{\mu}(\Lambda)$ is called a
\textit{$\mu$-twirl}.
A particular case of interest is when the measure $\mu$ is taken
as the unitarily-invariant Haar measure.
A unitary $2$-design has the property that sampling uniformly from
$\{U_1,\dots,U_K\}$ is operationally equivalent to sampling
from the Haar measure.
In other words, if $\mu$ is set to the uniform probability measure
on $\{U_1,\dots,U_K\}$ then, for any quantum channel $\Lambda$,
\be\label{eq:expect}
\mathbb{E}_{\mu}(\Lambda) = \mathbb{E}_{\Haar}(\Lambda).
\ee
This can be seen by considering the linear mappings $\Lambda$ on
$L(\mathbb{C}^D)$ of the form
\be\label{eq:def:axb}
\Lambda(X) = A X B,
\ee
where
$A, B \in L(\mathbb{C}^D)$.
Then the condition in Eq.~\ref{eqn:t-design} is equivalent to the
condition
\be\label{eq:axb}
\frac{1}{K} \sum_{k=1}^K U_k\dag A U_k X U_k\dag B U_k
= \int_{\mathcal{U}(D)} dU \; U\dag A U X U\dag B U
\ee
for all $A, X, B \in L(\mathbb{C}^D)$;
this equivalence can be seen explicitly by considering $A, X, B$ of
the form $\ket{i}\bbra{j}$.
Although not all quantum channels are of the form of Eq.~\ref{eq:def:axb},
they are convex combinations of mappings of this form.
Therefore, Eq.~\ref{eq:expect} follows by linearity.
Similarly, one can show that bilateral twirling \cite{BDSW,DLT02} and generating typical subsystem
purity \cite{ODP06} correspond to particular instances of Eq.~\ref{eqn:t-design} with t=2.

Our first contribution below is to extend the results of~\cite{BDSW,DLT02}
and prove the following.
\begin{theorem}\label{thm:clifford}
{\it The uniform distribution over the Clifford group on $n$ qubits
is a unitary $2$-design with $D=2^n$.}
\end{theorem}
\noindent
As is known, the Clifford group  $\mathcal{C}_n$ on $n$ qubits can
be implemented by quantum circuits of size $O(n^2)$
\cite{DLT02,Gottesman}. This general result allows us to immediately
deduce, for example, that the Clifford group gives a more efficient
solution to the protocol for generating generic entanglement given
in~\cite{ODP06}. Moreover, as described at the end of this paper,
this result leads to an efficient solution to the experimental
problem of estimating the average fidelity of a quantum process or a
quantum channel~\cite{EAZ05}.

Given the wide class of protocols which require unitary $2$-designs,
we show furthermore that a more efficient
implementation on $n$-qubits is possible if we consider
\emph{approximate} unitary $2$-designs.
We define these with respect to arbitrary linear superoperators
$\Lambda : L(\mathbb{C}^D) \rightarrow L(\mathbb{C}^D)$ (that is, $\Lambda$
need not be completely positive and trace preserving for this to make
sense).
Any such $\Lambda$ can be expressed in the form
$\Lambda(X) = \Tr_E ( A (X \otimes \openone_E) B )$, where $A$ and $B$
act on an extended Hilbert space $\mathcal{C}^D \otimes \mathcal{H}_E$.
A \emph{$\mu$-twirl} of the superoperator $\Lambda$ with respect to a
measure $\mu$ on a subset of $\mathcal{U}(D)$ is a mapping of the form
$\Lambda \mapsto \mathbb{E}_\mu(\Lambda)$ where
$\mathbb{E}_\mu(\Lambda)$ is the superoperator
%= K^{-1} \sum_k p_k U_k^\dagger \Lambda( U_k X U_k^\dagger) U_k $.
\begin{equation}
\mathbb{E}_\mu(\Lambda) : X \mapsto  \int_{\mathcal{U}(D)} d\mu(U) \;
U^\dagger \Lambda( U X U^\dagger) U.
\end{equation}
Note that, by linearity, unitary $2$-designs satisfy Eq.~\ref{eq:expect} for all
linear superoperators (i.e., not just for quantum channels).

Define an \emph{$\varepsilon$-approximate unitary $2$-design} in terms
of the diamond norm \cite{Kitaev} as a measure on a finite subset of
$\mathcal{U}(D)$ satisfying the property
\be
\label{eq:diamond} \| \mathbb{E}_\mu(\Lambda) -
\mathbb{E}_{\Haar}(\Lambda) \|_\diamond \leq \varepsilon
\| \Lambda \|_\diamond.
\ee
Note that in the interesting case where $\Lambda$ is a quantum channel,
$\| \Lambda \|_\diamond = 1$, hence the channel $\mathbb{E}_\mu(\Lambda)$
is a good approximation of the channel $\mathbb{E}_{\Haar}(\Lambda)$.

Our second contribution is to show that:
\begin{theorem}\label{thm:approx}{\it For all $\varepsilon > 0$, an
$\varepsilon$-approximate unitary $2$-design on $n$ qubits can be implemented
by in-place circuits of size $O(n \log 1/\varepsilon)$ and depth
$O(\log n \log 1/\varepsilon )$.}
\end{theorem}

Our third contribution is an application of this towards fidelity estimation:
\begin{theorem}\label{thm:appl}
{\it The average fidelity of a quantum channel $\Lambda$ acting on $n$
qubits, can be estimated to within $\delta > 0$ with error probability
$\varepsilon > 0$ at a cost of $O(\log 1/\varepsilon)$ evaluations of the channel conjugated by in-place circuits of size $O(n \log 1/\varepsilon)$ and depth $O(\log n \log 1/\varepsilon)$.}
\end{theorem}

% precise by noting that each
%side of Eq.~\ref{eq:axb} is the \textit{expectation} of a random
%variable $P_{(2,2)}(U)$ with respect to some measure on $U$. Thus,
%Eq.~\ref{eq:axb} can be expressed as $\EEE_{\mu}[P_{(2,2)}(U)] =
%\EEE_H[P_{(2,2)}(U)]$, where $\mu$ is a measure on
%$\{U_1,\dots,U_K\}$ and $H$ is the Haar measure. Now, define an
%\textit{$\varepsilon$-approximate unitary $2$-design} as a measure
%$\mu$ on a finite subset of $U(D)$ such that

%there exists a binary random variable $A$ that is correlated with
%$S$ so that $\EEE_{\mu}[P(U)|A=1] = \EEE_H[P(U)]$, and $\Pr[A=1] \ge
%1-\varepsilon$. It should be noted that our notion of an implementation
%here does not require $A$ to be explicitly sampled; in other words,
%although the failure probability must be bounded by $\varepsilon$,
%``detecting'' failures is not required.

%------------------------------------------------------------------------------&
\section{Exact construction}

\noindent
We prove Theorem~\ref{thm:clifford}, which implies that a unitary
$2$-design on $n$ qubits (dimension $D=2^n$) can be explicitly
constructed by in-place circuits of size $O(n^2)$.
Our approach is to construct a uniform probability
distribution on a subset of the Clifford group $\mathcal{C}_n$,
which defines a $\mathcal{C}_n$-twirl. It is sufficient to consider
linear mappings of the form $\Lambda(X) = A X B$, where $A$, $B \in
L(\mathbb{C}^D)$, and the results can be extended to arbitrary
linear superoperators by linearity.
%Let $\Lambda$ be any mapping of the form $\Lambda(X) = A X B$, where
%$A, B \in \mathbb{C}^{D \times D}$.
%Then applying a Clifford-twirl to $\Lambda$ is equivalent to
%applying a Haar-twirl to $\Lambda$.
Specifically,  we prove that, for all $X$,
\be
\frac{1}{|\mathcal{C}_n|}\sum_{U \in \mathcal{C}_n} U\dag A U X
U\dag B U  = \int_{U(D)} dU \; U\dag A U X U\dag B U. \nonumber
\ee
%\end{lemma}
As shown in Ref.~\cite{EAZ05}, the RHS can be expressed in the form,
\bes
& & \int_{U(D)} dU \; U\dag A U X U\dag B U   = \frac{\Tr(AB) \Tr(X)}{D}
\frac{\id}{D} \nonumber \\
\label{eqn:haar-twirl} & & + \frac{D \Tr(A)\Tr(B) -
\Tr(AB)}{D(D^2-1)}\left(X - \Tr(X) \frac{\id}{D}\right). \ees

To evaluate the LHS, we will make use of the fact $\mathcal{C}_n$ is
the normalizer of the generalized Pauli group $\mathcal{P}_n$
consisting of all $n$-fold tensor products of the one-qubit Pauli
operators $\{\id,X,Y,Z\}$. We denote the elements of $\mathcal{P}_n$
as $\{P_j\}_{j=1}^{D^2}$, where $P_1$ is the $n$-fold tensor product
of $\id$.
%\begin{lemma}\label{thm:pauli-twirl}
Applying a $\mathcal{P}_n$-twirl to the mapping $\Lambda(X) = A X B$
results in a mapping of the form $X \mapsto \sum_{k=1}^{D^2} r_k P_k
X P_k$ where $r_1 = \Tr(A)\Tr(B)/D^2$ and $\sum_{k=1}^{D^2}r_k =
\Tr(AB)/D$.
%\end{lemma}
This follows from noting that we can express $A = \sum_{a=1}^{D^2}
\alpha_a P_a$ and $B = \sum_{b=1}^{D^2} \beta_b P_b$. The resulting
operation maps $X$ to
\bes\label{InitPauli}
\lefteqn{1/D^2 \sum_{k=1}^{D^2} P_k A P_k X P_k B P_k} & & \nonumber \\
%& = & 1/D^2 \sum_{k=1}^{D^2} P_k \left(\sum_{a=1}^{D^2} \alpha_a
%P_a\right) P_k X P_k
%\left(\sum_{b=1}^{D^2} \beta_b P_b\right) P_k \nonumber \\
& = & 1/D^2 \sum_{a=1}^{D^2}\sum_{b=1}^{D^2} \alpha_a \beta_b
\left(\sum_{k=1}^{D^2}(-1)^{(k,a \oplus b)_{S_P}}\right)P_a X P_b
\nonumber \\
& = & \sum_{a=1}^{D^2} \alpha_a \beta_a P_a X P_a,
\ees
with the symplectic inner product $S_P$ on the index space (see
\cite{Dankert} for further details; these techniques are discussed
also in \cite{Cirac05}). Therefore, setting $r_k = \alpha_k
\beta_k$ leads to the above form.

%Using Lemma~\ref{thm:pauli-twirl}, we prove the following.
%\begin{lemma}\label{thm:clifford-twirl}

We can express each $U \in \mathcal{C}_n$ as $U = Q_j P_k$, where
$\{P_1,\dots,P_{D^2}\} = \mathcal{P}_n$ and
$\{Q_1,\dots,Q_{|\mathcal{P}_n|/|\mathcal{C}_n|}\}$ contains a
representative from each coset in $\mathcal{C}_n/\mathcal{P}_n$ (how
these representatives are chosen does not matter). Hence, after
twirling $\Lambda$ by $\mathcal{P}_n$, we then twirl with
$\{Q_1,\dots,Q_{|\mathcal{P}_n|/|\mathcal{C}_n|}\}$, where we
henceforth refer to the latter operation as a twirl by
$\mathcal{C}_n/\mathcal{P}_n$. The
$\mathcal{C}_n/\mathcal{P}_n$-twirl yields
\be
\frac{|\mathcal{P}_n|}{|\mathcal{C}_n|}
\sum_{j=1}^{|\mathcal{C}_n|/|\mathcal{P}_n|} \sum_{k=1}^{D^2} r_k
Q_j^\dagger P_k Q_j X Q_j^\dagger P_k Q_j.
\ee
Next we distinguish the identity element $P_1 = \id$ and make use of
the fact that conjugation under
%the Clifford group is the normalizer of the Pauli group and hence,
the Clifford group maps each non-identity Pauli element to every
other non-identity Pauli element with equal frequency. It follows
that the final state is
\bes\label{eqn:clif-twirl} \lefteqn{ r_1 X +
\frac{|\mathcal{P}_n|}{|\mathcal{C}_n|} \sum_{k=2}^{D^2} r_k
\sum_{j=1}^{|\mathcal{C}_n|/|\mathcal{P}_n|} Q_j^\dagger P_k Q_j X
Q_j^\dagger P_k Q_j
} & & \nonumber \\
& = & r_1 X + \frac{1}{D^2-1}\left(\sum_{k=2}^{D^2}r_k\right)
\sum_{l=2}^{D^2} P_l X P_l.
\ees Using the identity $\sum_{j=1}^{D^2} P_j X P_j = D \Tr(X)\id$,
it is straightforward to show that the right sides of
Eqs.~\Ref{eqn:clif-twirl} and~\Ref{eqn:haar-twirl} are equal.
%By linearity, we deduce the Theorem~\ref{thm:clifford}.

%\begin{proof}[Proof of Theorem~\ref{thm:approx}]
%{} From Lemma~\ref{thm:pauli-twirl},
%------------------------------------------------------------------------------&
\section{Approximate construction}

\begin{figure}%[ht!]
\fbox{
%\begin{minipage}{8.1cm}
\begin{minipage}{8.1cm}
\vspace*{3mm}
\noindent\textbf{Uniformization procedure:\hspace*{3.7cm}}
\begin{enumerate}
\item
$\mathcal{C}_1/\mathcal{P}_1$-twirl qubit $k$ for all $k \in
\{1,\dots,n\}$.
\item
Conjugate the first qubit by a random XOR. \newline
(This operation is defined in Figure~\ref{RandomXor} below.)
\item
$H$-conjugate the first qubit,
and $\mathcal{C}_1/\mathcal{P}_1$-twirl
qubit~$k$
for all $k \in \{2,\dots,n\}$.
\item
Conjugate the first qubit by a random XOR.
\item
$H$-conjugate the first qubit,
and $\mathcal{C}_1/\mathcal{P}_1$-twirl
qubit~$k$
for all $k \in \{2,\dots,n\}$.
\item
With probability $1/2$, $S$-conjugate the first qubit.
\item
Conjugate the first qubit by a random XOR.
\item
$\mathcal{C}_1/\mathcal{P}_1$-twirl the first qubit.
\end{enumerate}
\vspace*{2mm}
\end{minipage}
}
\caption{The uniformization procedure used in the approximate construction.}\label{fig:uniform}
\end{figure}

\noindent
We now prove Theorem~\ref{thm:approx}, that an $\varepsilon$-approximate
unitary $2$-design can be explicitly constructed in terms of circuits
that are in-place, of size $O(n \log 1/\varepsilon)$, and of depth
$O(\log n \log 1/\varepsilon)$.
More precisely, we describe a probabilistic construction that produces an $n$-qubit quantum circuit, generated according to a probability distribution
$(p_1,\dots,p_m)$ on a sequence of circuits $(C_1,\dots,C_m)$ with
the following property.
For any linear superoperator $\Lambda$ on $n$-qubits, the mapping
\be
\rho \mapsto \sum_{i=1}^m p_i C_i^{\dagger}\Lambda(C_i \rho C_i^{\dagger})C_i
\ee
is $\varepsilon$-close to (with respect to $\|\cdot\|_\diamond$) to the mapping
\be\label{PerfectTwirl}
\rho \mapsto
\int_{\mathcal{U}(2^n)} dU \; U^{\dagger}\Lambda(U \rho U^{\dagger})U.
\ee
Since we converting $\Lambda$ to the superoperator
\be
\sum_{i=1}^m p_i \hat{C_i}\dag \circ \Lambda \circ \hat{C_i},
\ee
we describe the probabilistic circuit construction as a series of simple
operations that are each conjugations performed on the channel.
%In other words, we start with the superoperator $\Lambda$ and perform a
%sequence of conjugations to this superoperator by elementary gates, ultimately
%leading to a superoperator that is $\varepsilon$-close to the mapping expressed by %Eq.~\ref{PerfectTwirl}.

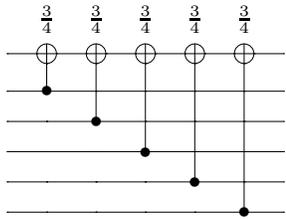
\begin{figure}%[ht!]
\begin{equation*}
\Qcircuit @C=1.2em @R=1em {
& \frac{3}{4} & \frac{3}{4} & \frac{3}{4} & \frac{3}{4} & \frac{3}{4} \\
& \targ   & \targ   & \targ   & \targ   & \targ   & \qw \\
& \ctrl{-1} & \qw   & \qw     & \qw     & \qw     & \qw \\
& \qw     & \ctrl{-2} & \qw   & \qw     & \qw     & \qw \\
& \qw     & \qw     & \ctrl{-3} & \qw   & \qw     & \qw \\
& \qw     & \qw     & \qw     & \ctrl{-4} & \qw   & \qw \\
& \qw     & \qw     & \qw     & \qw     & \ctrl{-5} & \qw \\
}
\end{equation*}
\caption{\textit{Conjugating the first qubit by a random XOR} is conjugation by a randomly generated circuit of the above form, where a numerical label associated with a gate indicates that it occurs with probability $3/4$ (with probability $1/4$ there is no gate).
That is, for each $k \in \{2,\dots,n\}$, with independent probability $3/4$, there is a CNOT gate with the first qubit as target and qubit $k$ as control.}\label{RandomXor}
\end{figure}

Our construction first applies a Pauli-twirl to the superoperator,
which consists of $O(n)$ gates and results in a superoperator that
is a linear combination of Pauli channels of the form $\rho \mapsto
P_a \rho P_a$. In order to convert an arbitrary Pauli channel into a
good approximation of a depolarizing channel, we shall add slightly more
than $O(n)$ further twirling operations to approximately uniformize
the probabilities associated with each $P_a$ for all $a \neq 1$.
The process consists of a series of repetitions of the procedure in
Figure~\ref{fig:uniform} (where the operation \textit{conjugating the
first qubit by a random XOR} is defined in Figure~\ref{RandomXor}).

A $\mathcal{C}_1/\mathcal{P}_1$-twirl of a qubit can be analyzed as
follows. Let $R = SH$, where $S = \ket{0}\bbra{0} + i\ket{1}\bbra{1}$,
and $H$ is the Hadamard transform. Select $i \in \{0, 1, 2\}$
uniformly and conjugate the register by $R^{i}$. This operation has
the property that, if it is applied to the identity channel $\id$,
it has no net effect; however, for a Pauli channel of the form $X$,
$Y$, or $Z$, this operation causes the channel to become a uniform
mixture of $X$, $Y$, and $Z$.

By Eq.~\ref{InitPauli}, the initial Pauli twirl results in a linear
combination of mappings of the form $\rho \mapsto P_a \rho P_a$. We
consider each term separately: start with a channel of the form
$\rho \mapsto P_a \rho P_a$, for some fixed $a \neq 1$, and apply
the above procedure. To analyze the result, we trace through the
effect of each of the eight steps of the uniformization procedure:

\begin{enumerate}

\item
For each $k$,
if component $k$ of $P_a$ is $\id$ then it remains $\id$, and if
component $k$ is $X$, $Y$, or $Z$ then it becomes a uniform mixture
of $X$, $Y$, and~$Z$.

\item
Call an execution of this procedure \textit{good} if, after Step 2,
the first component of the channel is $X$ or $Y$. This happens with
probability at least $1/2$, which can be seen by considering these
cases:

Case 1: For all $k \in \{2,\dots,n\}$, component $k$ of $P_a$ is
$\id$.
In this case, the CNOT gates have no effect, but since $a
\neq 1$, component $1$ of $P_a$ is \textit{not} $\id$.
Therefore, after the previous step, the first component of $P_a$
is uniformly distributed over $X$, $Y$, and $Z$.
Hence the first component of the channel is $X$ or $Y$ with
probability $2/3$.

Case 2: For some $k \in \{2,\dots,n\}$, component $k$ of $P_a$ is
not $\id$.
With probability $(2/3)(3/4) = 1/2$, component $k$ has
$X$ or $Y$ \textit{and} the CNOT gate is present.
This causes the first component to evolve as follows.
If it is $X$ or $\id$ then it becomes an equal mixture of $\id$
and $X$.
Also, if it is $Y$ or $Z$ then it becomes an equal mixture of $Y$
and $Z$.

In both of the above cases, the first component is $X$ or $Y$ with probability $1/2$.

\item
If the execution is good then
the first component
is $Y$ or $Z$.
For each $k \in \{2,\dots,n\}$, component $k$ is
either $\id$ or a uniform mixture of $X$, $Y$, and $Z$.

\item
If the execution is good then
for each
$k \in \{2,\dots,n\}$, component $k$ is $\id$ with independent
probability $1/4$, and some mixture of $X$, $Y$, $Z$ with
probability $3/4$.
To see why this is so, for each $k$, consider the effect of the
back-action of the CNOT gates in the following two cases separately.

Case 1: After the previous step, component $k$
%of the channel
is $\id$.
In this case,
it remains $\id$ with probability $1/4$,
and it becomes $Z$ with probability $3/4$.

Case 2: After the previous step, component $k$
is a uniform mixture of $X$, $Y$, and $Z$.
In this case, with probability $3/4$, the channel becomes a
uniform mixture of $Y$, $X$, and $\id$.
Hence the component becomes $\id$ with probability $(3/4)(1/3) = 1/4$.

\item
If the execution is good then, after this step, the first
component of the channel is $X$ or $Y$, and, for each
$k \in \{2,\dots,n\}$, component $k$ is independently a
uniform mixture of $\id$, $X$, $Y$, and $Z$.

\item
If the execution is good then, after this step, the first
component of the channel is a uniform mixture of $X$ and~$Y$.

\item
Call a good execution \textit{typical} if, after Step 6, there is at
least one component $k \in \{2,\dots,n\}$ that is not~$\id$.
The probability that a good execution is also typical is $1-(1/4)^{n-1}$.
If the execution is good and typical, the first component
of the channel is a uniform mixture of $\id$, $X$, $Y$, and
$Z$ (independent of the other components of the channel).

To see why this is so, consider the effect of any non-$\id$ component
$k \in \{2,\dots,n\}$.
Prior to the potential conjugation by CNOT, the first component is
uniformly distributed among $X$ and $Y$ and component $k$ is uniformly
distributed among $X$, $Y$, and $Z$.
Therefore, with probability $(2/3)(3/4) = 1/2$, the first component
becomes a uniform mixture of $I$ and $Z$.

\item
If the execution is good then, after this step, the first component
of the channel is: in a uniform mixture of $X$, $Y$, and $Z$ if
the execution is not typical; and a uniform mixture of $\id$, $X$, $Y$,
and $Z$ if it is typical.

\end{enumerate}

For executions that are both good and typical, there are $4(4^{n-1}-1)$
possible Pauli channels that can result (namely, those that are
not $\id$ in at least one of the components from $2$ through $n$).
Conditional on the execution being good, each of these cases arises with probability $1/4^n$.
For executions that are good but not typical, there are three possible
outcomes (namely, all channels that are $X$, $Y$, or $Z$ in the first
component and $\id$ in components $2$ through~$n$).
Conditional on the execution being good, each of these three cases arises
with probability ${1/(3 \cdot 4^{n-1})}$.
The resulting probability distribution on Pauli channels can be expressed
as a convex combination of these two distributions:
(a) the uniform distribution of all non-trivial Pauli channels; and
(b) the uniform distribution on the three non-typical Pauli channels.
Distribution (a) occurs with probability $(1/2)(1-(1/4)^n)$, and
distribution (b) occurs with probability $(1/2)(1+(1/4)^n)$.
Note that distribution (a) corresponds to a perfect 2-design.
Repeating the procedure $O(\log 1/\varepsilon)$ times, we can
increase probability weighting associated with (a) from
$(1/2)(1-(1/4)^n)$ to $1 - \varepsilon/2$ (the perfect Pauli channel
need only arise in one of the repetitions).
%Conditional on the execution being good, it can be shown that the
%non-identity Pauli operators of the channel are distributed as a
%convex combination of these two distributions: the uniform
%distribution on all non-identity Pauli operators; and the uniform
%distribution on the three non-identity Pauli operators that are
%$\id$ in their last $n-1$ components.
%Moreover, the probability of
%the second distribution occurring is less than $1/2$. It follows
%that the resulting channel is a perfect satisfies the definition of
%an $\varepsilon$-approximate unitary 2-design.
Each execution of the uniformization procedure consists of $O(n)$ gates,
that can be implemented in $O(\log n)$ depth---the only nontrivial part
is the conjugations by a random XOR, whose log-depth implementation is
based on the construction in Fig.~\ref{LogCircuit}.
%\end{proof}

\begin{figure}
\begin{equation*}
\Qcircuit @C=1.2em @R=1em {
& \targ   & \targ   & \targ   & \targ   & \targ   & \targ   & \targ & \qw \\
& \ctrl{-1} & \qw   & \qw     & \qw     & \qw     & \qw     & \qw & \qw \\
& \qw     & \ctrl{-2} & \qw   & \qw     & \qw     & \qw     & \qw & \qw \\
& \qw     & \qw     & \ctrl{-3} & \qw   & \qw     & \qw     & \qw & \qw \\
& \qw     & \qw     & \qw     & \ctrl{-4} & \qw   & \qw     & \qw & \qw \\
& \qw     & \qw     & \qw     & \qw     & \ctrl{-5} & \qw   & \qw & \qw \\
& \qw     & \qw     & \qw     & \qw     & \qw     & \ctrl{-6} & \qw & \qw \\
& \qw     & \qw     & \qw     & \qw     & \qw     & \qw     & \ctrl{-7} & \qw \\
\\
& & & & \mbox{(A)} \\
\\
}
\end{equation*}
\begin{equation*}
\Qcircuit @C=1.2em @R=1em {
& \targ      & \targ     & \targ     & \qw       & \qw       & \qw \\
& \ctrl{-1}  & \qw       & \qw       & \qw       & \qw       & \qw \\
& \targ      & \ctrl{-2} & \qw       & \qw       & \targ     & \qw \\
& \ctrl{-1}  & \qw       & \qw       & \qw       & \ctrl{-1} & \qw \\
& \targ      & \targ     & \ctrl{-4} & \targ     & \targ     & \qw \\
& \ctrl{-1}  & \qw       & \qw       & \qw       & \ctrl{-1} & \qw \\
& \targ      & \ctrl{-2} & \qw       & \ctrl{-2} & \targ     & \qw \\
& \ctrl{-1}  & \qw       & \qw       & \qw       & \ctrl{-1} & \qw \\
\\
& & & \mbox{(B)} \\
\\
}
\end{equation*}
\caption{Part (A) shows a circuit consisting of $n=7$ CNOT gates with
common target; part (B) shows as equivalent circuit (based on a binary
tree of additions modulo 2) whose depth is bounded by $2\log n$ and
size is bounded $2n$.}\label{LogCircuit}
\end{figure}

The net result of this construction can be viewed a mixture of two
mixed-unitary operations, one of which is a perfect two-design.
The perfect $2$-design occurs with probability at least $1 - \varepsilon/2$
and the other operation occurs with probability at most $\varepsilon/2$.
Therefore the construction yields a linear superoperator of the form
\begin{equation}
(1 - \varepsilon/2)\, \mathbb{E}_{\Haar}(\Lambda)
+ (\varepsilon/2)\, \mathbb{E}_{\nu}(\Lambda),
\end{equation}
for some probability measure $\nu$ on $\mathcal{U}(2^n)$.
The diamond norm distance between this operation and $\mathbb{E}_{\Haar}(\Lambda)$ is bounded by
\begin{eqnarray}
\| (1 - \varepsilon/2)\, \mathbb{E}_{\Haar}(\Lambda) & + &
(\varepsilon/2)\, \mathbb{E}_{\nu}(\Lambda) - \mathbb{E}_{\Haar}(\Lambda)\|_\diamond \nonumber \\
& \le & (\varepsilon/2)\, \| \mathbb{E}_{\Haar}(\Lambda) \|_\diamond
+ (\varepsilon/2)\, \|\mathbb{E}_\nu(\Lambda)\|_\diamond \nonumber \\
& = & \varepsilon \| \Lambda \|_\diamond,
\end{eqnarray}
where we have made use of the fact that $\|\cdot\|_\diamond$ is
invariant under twirling:
$\|\mathbb{E}_\mu(\Lambda) \|_\diamond = \| \Lambda \|_\diamond$.
It follows that the construction produces an $\varepsilon$-approximate
$2$-design.

%[Leftover:]
%We are interested in two special cases. First,
%we take $\Lambda$ to have the form $\Lambda(X) = A X B$, , where
%$A$, $B \in L(\mathbb{C}^D)$, which corresponds to the argument in
%the exact 2-design condition given by Eq.~\ref{eq:axb}. In this case
%our approximate construction satisfies
%\be
%|| (1 - \tilde{\varepsilon}) \mathbb{E}_{\Haar}(\Lambda) +
%\tilde{\varepsilon} \;  \mathbb{E}_{\nu}(\Lambda) -
%\mathbb{E}_{\Haar}(\Lambda)||_\diamond \leq 2 \tilde{\varepsilon} ||A ||\cdot
%||B ||
%\ee
%Second, we are interested in the physically relevant case that
%$\Lambda$ corresponds to a quantum channel. From the Stinespring
%dilation theorem we known that any quantum channel admits a
%description where $A$ and $B$ are unitary operators acting on the
%extended space $\mathcal{C}^D \otimes \mathcal{H}_E$. In this case
%our approximate construction satisfies
%\be
%|| (1 - \tilde{\varepsilon}) \mathbb{E}_{\Haar}(\Lambda) +
%\tilde{\varepsilon} \; \mathbb{E}_{\nu}(\Lambda) -
%\mathbb{E}_{\Haar}(\Lambda)||_\diamond \leq 2 \tilde{\varepsilon}
%\ee
%where we have made use of the fact that $|| \mathbb{E}(\Lambda)
%||_\diamond =1$ for any twirled quantum channel.

%------------------------------------------------------------------------------&
\section{Application to fidelity estimation}

\noindent
We now turn to a discussion of the experimental problem of fidelity
estimation for which the above unitary 2-design constructions lead
to an efficient, scalable protocol that is accessible with current
experimental techniques on systems of a few qubits. Consider the
Haar-averaged fidelity \cite{Nielsen02,EAZ05}
\bes\label{eqn:ave-fidelity}
\< F \> & \equiv & \int_{\mathcal{U}(D)} dU \; \Tr[ U | 0 \> \< 0 | U^\dagger
\Lambda( U | 0 \> \< 0 | U^\dagger] \nonumber \\
& = & \sum_k \frac{|\Tr(A_k)|^2 + D}{D^2 +D}.
\ees
of a quantum operation $\Lambda(\rho) = \sum_k A_k \rho
A_k^\dagger$. The Haar-averaged fidelity can be related to two
standard fidelity benchmarks: the entanglement-fidelity $F_e$, which
has been proposed as means of characterizing the noise strength in a
physical quantum channel $\Lambda$ \cite{Sch96}, and the
gate-fidelity $F_g$, which has been used to characterize the quality
of quantum memory \cite{Boulant} or of an implementation of a target
unitary $U_g$ on a noisy quantum processing device
\cite{Fortunato02,Yaakov}. In the latter scenario we imagine the
implementation of a gate sequence $U_g$ followed immediately by its
inverse $U_g^\dagger$, and make the identification $\Lambda(\rho) =
U_g^\dagger \E( U_g \rho U_g^\dagger) U_g$, where the map $\E(\rho)$
represents the noise accumulated over the course of implementing
$U_g^\dagger U_g$. Then, using the results of
Ref.~\cite{Sch96,Fortunato02,Nielsen02,EAZ05}, we find the following
relationship between the Haar-average fidelity and the previously
proposed gate-fidelity and entanglement-fidelity,
\be
\< F \>  = \frac{D F_g + 1}{D+1} = \frac{D F_e + 1}{D+1}.
\ee
We emphasize that this relationship holds in the context where $F_g$
and $F_e$ are understood to characterize errors under the composed
sequence $U_g^\dagger U_g$, rather than errors under $U_g$ itself.

There are two experimental approaches to estimating $F_e$ and $F_g$
for a given quantum channel. The first is based on ancilla-assisted
process tomography, or some variant such as direct characterization,
both of which require creating an entangled state of 2n qubits, the
first $n$ of which are subjected to the unknown transformation and
the remaining $n$ of which are ancilla qubits that are subject to
the identity channel \cite{AAPTNC-QPT}, followed by joint
measurements on the final $2n$ qubit state. A significant
disadvantage of this approach is the requirement of $n$ noise-free
ancilla qubits, as well as the requirement of joint operations on
$2n$, rather than $n$, qubits. The second approach is standard
process tomography, which suffers from the requirement of a number
of experiments that grows exponentially with $n = \log_2D$
\cite{AAPTNC-QPT,Nielsen02}.

However, as described in Ref.~\cite{EAZ05}, we can estimate $\< F
\>$ directly by the following protocol: apply a random unitary
operator $U$ to the initial state $|0\>$, followed by the quantum
operation $\Lambda$, and then apply $U^\dagger$ to the output state.
Then from Eq.~\Ref{eqn:ave-fidelity} we see that $\< F \>$ can be
estimated by repeating this procedure with $U$ sampled randomly from
the Haar measure in each experiment. For an arbitrary, but fixed,
average fidelity $0 \leq \< F \> \leq 1$, the number of experiments
required to estimate $\<F\> $ to precision $\delta > 1/4^n$ is
independent of the dimension $D$. A serious limitation of the
approach of Ref.~\cite{EAZ05} is that the implementation of a random
unitary requires exponential resources. However, given that $F$ is a
polynomial function of homogeneous degree (2,2), it follows that we
can estimate $\< F\>$ by sampling from any unitary 2-design instead
of the Haar-random unitary operators presumed in Ref.~\cite{EAZ05}.
Hence the results of this paper, and in particular the
$\varepsilon$-approximate unitary 2-design described above, imply
that each experiment requires only $O(n\log(1/\varepsilon))$ gates.
Hence the fidelity $\<F\>$, and equivalently $F_g$ and $F_e$, may be
estimated by an efficient experimental protocol that can be applied
with existing levels of quantum control in systems of a few qubits.
 We remark that, after the original submission of this work, a
randomization approach has been developed \cite{Emerson07} which
offers an improvement over the resource requirements discussed above
for the task of fidelity estimation. However, the randomization
approach of this paper is strictly stronger than that of
Ref.~\cite{Emerson07}.

It remains an interesting open question whether an arbitrary quantum
randomization algorithm can be reduced to a $t$-design condition, and
hence classified within this framework. This would provide further
motivation to generalize the methods of this paper to obtain unitary
and state $t$-designs for $t > 2$. The alternate definition proposed
in Note \cite{endnote2} might be a good starting point for research in
this direction.

%------------------------------------------------------------------------------&
\begin{acknowledgments}
We thank Robin Blume-Kohout, David Cory, Daniel Gottesman, Debbie
Leung, Pranab Sen, and John Watrous for helpful discussions and
Markus Grassl for pointing out an omission in an earlier version of
this work. This research was supported by Canada's NSERC, MITACS,
and CIFAR, and the U.S. ARO.
\end{acknowledgments}

%------------------------------------------------------------------------------&


\begin{thebibliography}{99}
\bibitem{endnote2}
An equivalent definition of a unitary $t$-design is such that
$\sum_{k=1}^K D^\ii(U_k) = 0$ for all non-trivial irreducible
representations $D^\ii$ contained in the tensor power $V^{\otimes t}
\otimes \overline{V}^{\otimes t}$ of the fundamental representation
$V$ and its conjugate. This characterization is especially relevant
when analyzing the efficiency of ``random circuit'' constructions
for generating pseudo-random sets of unitaries \cite{ELL}.
\end{thebibliography}
\end{document}